\documentclass{PoS}
\usepackage{amssymb,amsmath}
\usepackage{booktabs}
\usepackage{slashed}
\usepackage[pdftex]{graphicx}

\newcommand{\mysmall}[1]{\scriptscriptstyle \rm #1}
\newcommand{ \deltaw}{\delta_{\mysmall{W}}}
\newcommand{ \mw}{M_{\mysmall{W}}}
\newcommand{ \rw}{r_{\mysmall{W}}}
\newcommand{ \gw}{g_{\mysmall{W}}}
\newcommand{\ymin}{y_0}

\newcommand{\Lidue}{\textup{Li}_2}
\newcommand{\gmt}{$g$$-$$2$ }

\newcommand{ \cw}{\cos \theta_{\mysmall W}}
\newcommand{ \sw}{\sin \theta_{\mysmall W}}

\newcommand{ \Clb}{C_{lB}}
\newcommand{ \Clw}{C_{lW}}
\renewcommand{\Re}{{\rm Re}}
\renewcommand{\Im}{{\rm Im}}

\title{Precision tests via radiative $\mu$ and $\tau$ leptonic decays}

\ShortTitle{Precision tests via radiative $\mu$ and $\tau$ leptonic decays}

\author{Matteo~Fael\\
        Albert Einstein Center for Fundamental Physics\\
	Institute for Theoretical Physics, University of Bern \\ CH-3012 Bern, Switzerland\\
        E-mail: \email{fael@itp.unibe.ch}}

\author{\speaker{Massimo~Passera}\\
        Istituto Nazionale Fisica Nucleare, Sezione di Padova \\ I-35131 Padova, Italy\\
        E-mail: \email{passera@pd.infn.it}}

\abstract{The branching fractions of radiative leptonic $\tau$ decays $(\tau \to l \nu \bar{\nu} \gamma$, $l=e,\mu)$ were recently measured by the \textsc{Babar} collaboration with a relative error of about 3\%. The measurement of the branching ratio $\mathcal{B} (\tau \to e \bar{\nu}  \nu \gamma)$, for a minimum photon energy of 10~MeV in the $\tau$ rest frame, differs from our recent SM prediction by 3.5 standard deviations, whereas our result agrees with \textsc{Babar}'s value for $\mathcal{B} (\tau \to \mu \bar{\nu} \nu \gamma)$. Our predictions also agree with the measurement of $\mathcal{B} (\mu \to e \bar{\nu} \nu \gamma)$ by the \textsc{Meg} collaboration. We also report on a recent proposal to test the $\tau$ dipole moments via precise measurements of radiative leptonic $\tau$ decays at high-luminosity $B$ factories.}

\FullConference{12th International Symposium on Radiative Corrections (Radcor 2015) and LoopFest XIV (Radiative Corrections for the LHC and Future Colliders) \\ 15-19 June, 2015\\UCLA Department of Physics \& Astronomy, Los Angeles, USA}

\begin{document}

\section{Introduction}

\label{intro}

Muon and $\tau$ leptonic decays have been among the most powerful tools to study the Lorentz structure of weak interactions. Their precise theoretical formulation in terms of Bouchiat-Michel-Kinoshita-Sirlin parameters~\cite{Michel:1949qe,Bouchiat:1957zz,KinoshitaSirlin:1957} places them in a unique position to investigate possible contributions beyond the $V$--$A$ coupling of the Standard Model (SM). Radiative $\mu$ and $\tau$ leptonic decays, where an inner bremsstrahlung photon is emitted and detected, can be predicted with very high precision and provide independent and complementary tests~\cite{Eichenberger:1984gi,Fetscher:1993ki}. Radiative $\mu$ and $\tau$ leptonic decays also constitute an important source of background for experiments searching for charged lepton flavor violating decays, such as $\mu^\pm \to e^\pm \gamma$, $\tau^\pm \to l^\pm \gamma$ ($l=e,\mu$), and even $\mu^\pm \to  e^\pm (e^+ e^-)$ and $\tau^\pm \to l^{\pm} (e^+ e^-)$, because of the internal conversion of photons to electron-positron pairs~\cite{Adam:2013gfn,Baldini:2013ke,Berger:2014vba,Flores-Tlalpa:2015vga}.

Recently the \textsc{Babar} collaboration measured the $\tau \to l \gamma \nu \bar{\nu}$ branching ratios for a minimum photon energy $\omega_0=10$~MeV in the $\tau$ rest frame~\cite{Lees:2015gea,OberhofPhDThesis}. These measurements, with a relative error of about 3\%, must be compared with the SM predictions of the branching fractions at next-to-leading order (NLO). Indeed these radiative corrections, recently computed in~\cite{Fael:2015gua}, are not protected from mass singularities by the Kinoshita-Lee-Nauenberg (KLN) theorem~\cite{Kinoshita:1958ru,Kinoshita:1962ur,Lee:1964is} and are of relative order $(\alpha/\pi) \ln(m_l/m_\tau) \ln(\omega_0/m_\tau)$, corresponding to a large $10\%$ correction for $l=e$, and $3\%$ for $l=\mu$. Radiative muon decays were measured long ago in~\cite{Crittenden:1959hm}, and new results were presented recently by the \textsc{Meg}~\cite{Adam:2013gfn} and \textsc{Pibeta}~\cite{Pocanic:2015rrw} collaborations.

Precise data on radiative $\tau$ leptonic decays also offer the opportunity to probe the $\tau$ anomalous magnetic moment ($g$$-$$2$) and electric dipole moment (EDM). The short lifetime of the $\tau$ has so far prevented the direct measurement of its \gmt via the $\tau$ spin precession in a magnetic field (like in the electron and muon \gmt experiments) and the present bound is only of $O(10^{-2})$, more than an order of magnitude larger than the leading contribution $\alpha/(2\pi) \approx 0.001$. While experiments attempted the extraction of indirect bounds from $\tau$ pair production and decays by comparing sufficiently precise data with the SM predictions, in~\cite{Fael:2013ij, FaelThesis,nextfael} we proposed the study of the $\tau$ electromagnetic dipole moments via its radiative leptonic $\tau$ decays by means of an effective Lagrangian approach.

In section~\ref{sec:raddecay} and~\ref{sec:BR} we present the SM predictions for differential decay rates and branching ratios of radiative $\mu$ and $\tau$ leptonic decays, and compare them with the experimental results. In section~\ref{sec:ad} we review the current status of the $\tau$ dipole moments and report on the sensitivities that can be expected at the upcoming Belle II experiment. Conclusions are drawn in section~\ref{sec:c}.

\section{{\boldmath  Radiative $\mu$ and $\tau$ leptonic decays}: differential rates}\label{sec:raddecay}

The SM prediction, at NLO, for the differential rate of the radiative leptonic decays 
\begin{align}
  \mu^\pm &\to e^\pm \nu \, \bar{\nu} \, \gamma, \label{eqn:muraddecay} \\
   \tau^\pm &\to l^\pm \, \nu \, \bar{\nu} \, \gamma, \label{eqn:tauraddecays}
\end{align}
with $l=e$ or $\mu$, of a polarized $\mu^\pm$ or $\tau^\pm$ in their rest frame is
\begin{equation}
   \frac{d^6 \Gamma^\pm \left(\ymin\right) }{dx \, dy \, d\Omega_l\, d\Omega_\gamma}  =
	\frac{\alpha \, G_F^2 M^5} {(4 \pi)^6} 
	\frac{x \beta_l}{1+ \deltaw} \,
	\biggl[
	G
	\, \mp \, x \beta_l \, \hat{n} \cdot \hat{p}_l  \, J 
	\, \mp \, y \, \hat{n} \cdot \hat{p}_\gamma \, K 
        \, + \, x y \beta_l \, \hat{n} \cdot \left(\hat{p}_l \times \hat{p}_\gamma \right) L
	\biggr],
  \label{eqn:radiativedecayrateNLO}
\end{equation}
where 
$G_F=1.166 \, 378 \, 7(6) \times10^{-5}$ GeV$^{-2}$~\cite{PDG2014} 
is the Fermi constant determined by the muon lifetime, and
$\alpha = 1/137.035\,999\,157\,(33)$
is the fine-structure constant~\cite{AHKN}. Calling $m$ and $M$ the masses of the final and initial charged leptons (neutrinos and antineutrinos are considered massless) we define $r=m/M$ and $\rw=M/\mw$, where $\mw$ is the $W$-boson mass; $p$ and $n=(0,\hat{n})$ are the four-momentum and polarization vector of the initial $\tau$ or muon, with $n^2=-1$ and $n \cdot p = 0$. Also, $x = 2E_l/M$, $y = 2\omega/M$ and $\beta_l \equiv |\vec{p}_l|/E_l=\sqrt{1-4r^2/x^2}$, where  $p_l = (E_l,\vec{p}_l)$ and $p_\gamma = (\omega,\vec{p}_\gamma)$ are the four-momenta of the final charged lepton and photon, respectively. The final charged lepton and photon are emitted at solid angles $\Omega_l$ and $\Omega_{\gamma}$, with normalized three-momenta $\hat{p}_l$ and $\hat{p}_\gamma$, and  $c$ is the cosine of the angle between $\hat{p}_l$ and $\hat{p}_\gamma$. The term $ \deltaw =1.04 \times 10^{-6}$ is the tree-level correction to muon decay induced by the $W$-boson propagator~\cite{Ferroglia:2013dga,Fael:2013pja}. Equation~\eqref{eqn:radiativedecayrateNLO} includes the possible emission of an additional soft photon with normalized energy $y'$ lower than the detection threshold $\ymin$ (with $\ymin \ll 1$): $y'<\ymin<y$. 
The function $G (x,y,c,\ymin)$ and, analogously, $J$ and $K$, are given by
\begin{equation}
  G \, (x,y,c,\ymin) = \\
 = \frac{4}{3 y z^2} 
  \bigg[ 
     g_0 (x,y,z)  
     + \rw^2 \, \gw  (x,y,z)
     + \frac{\alpha}{\pi} \, g_{\mysmall NLO} (x,y,z,\ymin) 
   \bigg],
  \label{eqn:GNLO}
\end{equation}
where $z=xy(1-c\beta_l)/2$; the LO function $g_0 (x,y,z)$, computed in~\cite{Fronsdal:1959zzb,EcksteinPratt,Kuno:1999jp}, arises from the pure Fermi $V$--$A$ interaction, whereas $\gw(x,y,z)$ is the LO contribution of the $W$-boson propagator derived in~\cite{Fael:2013pja}. The NLO term $g_{\mysmall NLO} (x,y,z,\ymin)$ is the sum of the virtual and soft bremsstrahlung contributions calculated in~\cite{Fael:2015gua} (see also~\cite{Fischer:1994pn,Arbuzov:2004wr}). 
The function $L(x,y,z)$, appearing in front of the product $\hat{n} \cdot \left(\hat{p}_l \times \hat{p}_\gamma \right)$, does not depend on $\ymin$; it is only induced by the loop corrections and is therefore of $\mathcal{O}(\alpha/\pi)$. The (lengthy) explicit expressions of $G,J,K$ and $L$ are provided in~\cite{Fael:2015gua}. 
If the initial $\mu^\pm$ or $\tau^{\pm}$ are not polarized, Eq.~\eqref{eqn:radiativedecayrateNLO} simplifies to
\begin{equation}	
\frac{d^3 \Gamma^\pm  \left(\ymin\right) }{dx \, dc \, dy}  =
	\frac{\,\alpha G_F^2 M^5} {(4 \pi)^6} \frac{x \beta_l}{1+ \deltaw}  \,\, 8 \pi^2 \, G \, (x,y,c,\ymin).
\label{eq:radiativedecayrateunpolarizedNLO}
\end{equation}

\section{{\boldmath  Radiative $\mu$ and $\tau$ leptonic decays}: branching ratios}\label{sec:BR}

The analytic integration of the LO part of the differential rate~\eqref{eq:radiativedecayrateunpolarizedNLO} over the allowed kinematic ranges for a minimum photon energy $\ymin=2\omega_0/M$ gives~\cite{EcksteinPratt,KSPRL1959}
\begin{align}
   \Gamma_{0}  \left( \ymin \right) & = 
   \, \frac{G_F^2 M^5}{192 \pi^3}
   \frac{\alpha}{3\pi} \, \bigg[
   3 \, \Lidue (\ymin)  - \frac{\pi^2}{2} + \left( \ln r +\frac{17}{12} \right) \left( 6 \ln \ymin +6 \bar{y}_0 + \bar{y}_0^4 \right)   \, +\\
  & \,\,\,\,\,\,\,\,\,\,\,\, + \frac{1}{48} \left(125+45 \ymin -33 \ymin^2 +7 \ymin^3 \right) \bar{y}_0 
    \, - \, \frac{1}{2} \left( 6+\bar{y}_0^3 \right) \bar{y} _0  \ln \bar{y}_0 \bigg],
 \label{eqn:totalLOr}
 \end{align}
where $\bar{y}_0 = 1-\ymin$ and the dilogarithm is defined by
$\Lidue(z) = -\int_0^z \!dt \,\frac{\ln(1-t)}{t}.$
Terms depending on the mass ratio $r$ have been neglected in the expression for $\Gamma_{0}(\ymin)$, with the obvious exception of the logarithmic contribution which diverges in the limit $r\to0$. However, terms in the integrand $g_{0} (x,y,c)$ (see \eqref{eqn:GNLO}) proportional to $r^2$ were not neglected when performing the integral to obtain~\eqref{eqn:totalLOr}, as they lead to terms of $\mathcal{O}(1)$ in the integrated result $\Gamma_{0}(\ymin)$. This feature was first pointed out in~\cite{Lee:1964is}. 
We also note that the presence of the mass singularity $\ln r$ in the integrated decay rate $\Gamma_{0}  \left( \ymin \right)$ does not contradict the KLN theorem, which applies only to total decay rates~\cite{Kinoshita:1958ru, Kinoshita:1962ur, Lee:1964is}.
The tiny corrections induced by the $W$-boson propagator were neglected in~eq.~\eqref{eqn:totalLOr}.

\begin{table}[htb]
  \def\arraystretch{1.1}
   \centering
\small
\setlength{\tabcolsep}{4pt}
   \begin{tabular}{lrrrr}
     \toprule &
	$\tau \to e \bar{\nu} \nu \gamma$~~\cite{Lees:2015gea}	&
	$\tau \to \mu \bar{\nu} \nu \gamma$~~\cite{Lees:2015gea}	&
	$\mu \to e \nu \bar{\nu}\gamma$~~\cite{Crittenden:1959hm} 	&
	$\mu \to e \nu \bar{\nu}\gamma$~~\cite{Adam:2013gfn} 	\\
      	\hline
	$\mathcal{B}_{\scriptscriptstyle \rm LO}$ 				&	
	$ 1.834 \cdot 10^{-2}$ &
      	$ 3.663 \cdot 10^{-3}$ &
 	$ 1.308 \cdot 10^{-2}$ &
	$ 6.204 \cdot 10^{-8}$ \\
	$\mathcal{B}_{\scriptscriptstyle \rm NLO}^{\rm Inc}$ 		&
      	$-1.06  \, (1)_n (10)_N \cdot 10^{-3}$ &     
      	$-5.8  \, (1)_n (2)_N \cdot 10^{-5}$ &     
      	$-1.91  \, (5)_n (6)_N \cdot 10^{-4}$ &     
      	$-3.61  \, (8)_n (21)_N \cdot 10^{-9}$ \\     
	$\mathcal{B}_{\scriptscriptstyle \rm NLO}^{\rm Exc}$	&
	$-1.89  \, (1)_{n} (19)_N \cdot 10^{-3}$ &     
      	$-9.1  \, (1)_{n} (3)_N \cdot 10^{-5}$ &     
      	$-2.25  \, (5)_{n}  (7)_N \cdot 10^{-4}$ &     
      	$-3.61  \, (8)_n (21)_N \cdot 10^{-9}$ \\     
	$\mathcal{B}^{\rm Inc}$							&
      	$ 1.728  \, (10)_{\rm th} (3)_{\rm \tau} \cdot 10^{-2}$ &      
      	$ 3.605  \, (2)_{\rm th} (6)_{\rm \tau} \cdot 10^{-3}$ &      
      	$ 1.289  \, (1)_{\rm th} \cdot 10^{-2}$ &      
      	$ 5.84  \, (2)_{\rm th} \cdot 10^{-8}$ \\      
	$\mathcal{B}^{\rm Exc}$							&
      	$ 1.645  \, (19)_{\rm th} (3)_{\rm \tau} \cdot 10^{-2}$ &      
      	$ 3.572  \, (3)_{\rm th} (6)_{\rm \tau} \cdot 10^{-3}$ &      
      	$ 1.286  \, (1)_{\rm th} \cdot 10^{-2}$ &     
      	$ 5.84  \, (2)_{\rm th} \cdot 10^{-8}$ \\      
 	$\mathcal{B}_{\scriptscriptstyle \rm EXP}$				&
      	$ 1.847  \, (15)_{\rm st} (52)_{\rm sy} \cdot 10^{-2}$ &      
      	$ 3.69  \, (3)_{\rm st} (10)_{\rm sy} \cdot 10^{-3}$ &      
      	$ 1.4  \, (4) \cdot 10^{-2}$ &     
      	$ 6.03  \, (14)_{\rm st} (53)_{\rm sy} \cdot 10^{-8}$ \\      
\bottomrule
\end{tabular}
\caption{Branching ratios of radiative $\mu$ and $\tau$ leptonic decays. The minimum photon energy $\omega_0$ is 10~MeV, except for the last column, where $\omega_0=40$~MeV and $E_e^{\rm min}=45$~MeV. Inclusive and exclusive ($\mathcal{B}^{\rm Inc/Exc}$) predictions are separated into LO contributions ($\mathcal{B}_{\scriptscriptstyle \rm LO}$) and NLO corrections ($\mathcal{B}_{\scriptscriptstyle \rm NLO}^{\rm Inc/Exc}$). Uncertainties were estimated for uncomputed NNLO corrections ($N$), numerical errors ($n$), and the experimental errors of the lifetimes ($\tau$). The first two types of errors were combined to provide the total theoretical uncertainty (th). The last line reports the experimental measurements of Refs.~\cite{Lees:2015gea,Crittenden:1959hm,Adam:2013gfn}. 
\label{tab:BR}}
\end{table}

If we multiply the analytic result for $\Gamma_{0} \left( \ymin \right)$ in eq.~\eqref{eqn:totalLOr} by the lifetimes $\tau_{\mu,\tau}$ with a threshold $\omega_0 = 10$~MeV we obtain the following LO predictions for the branching ratios:
$1.83 \times 10^{-2}~(\tau \to e \bar{\nu} \nu \gamma)$, 
$3.58 \times 10^{-3}~(\tau \to \mu \bar{\nu} \nu \gamma)$, and
$1.31 \times10^{-2}~(\mu \to e \nu \bar{\nu}\gamma)$.
These values are in good agreement with the results $\mathcal{B}_{\scriptscriptstyle \rm LO}$ reported in table~\ref{tab:BR}, obtained integrating numerically the LO part of the differential rate~\eqref{eq:radiativedecayrateunpolarizedNLO}, with the exception of the $\tau \to \mu \bar{\nu} \nu \gamma$ value; this difference is due to the terms neglected in the analytic result~\eqref{eqn:totalLOr}.

At NLO, which allows for double photon emission, the branching ratios of the radiative decays (\ref{eqn:muraddecay},\ref{eqn:tauraddecays}) can be distinguished in two types:
\begin{itemize}
\item "Inclusive" measurements of the branching ratios, $\mathcal{B}^{\rm Inc} \left( y_0 \right)$, where there is {\em at least one} photon in the final state with energy higher than $\ymin$; 
\item "Exclusive" measurements of the branching ratios, $\mathcal{B}^{\rm Exc} \left( y_0 \right)$, where there is {\em one, and only one}, photon in the final state with energy larger than the detection threshold $y_0$.
\end{itemize}
Exclusive and inclusive branching ratios for the radiative decays (\ref{eqn:muraddecay},\ref{eqn:tauraddecays}) were computed in~\cite{Fael:2015gua} for a threshold $\omega_0 = \ymin \, (M/2) = 10$~MeV, and are reported in table~\ref{tab:BR}.
Uncertainties were estimated for uncomputed NNLO corrections, numerical errors, and the experimental errors of the lifetimes.
For $\omega_0 = 10$~MeV, the former were estimated to be  
$\delta \mathcal{B}^{\rm Exc/Inc}_{\mysmall NLO} \! \sim (\alpha/\pi) \ln r \ln (\omega_0/M) \, \mathcal{B}_{\mysmall NLO}^{\rm Exc/Inc} \!;$
they are about
10\%, 3\% and 3\%
for $\tau \to e \bar{\nu} \nu \gamma$, $\tau \to \mu \bar{\nu} \nu \gamma$ and $\mu \to e \bar{\nu} \nu \gamma$, respectively (they appear with the subscript "$N$" in table~\ref{tab:BR}). 
Numerical errors, labeled by the subscript "$n$", are smaller than those induced by missing radiative corrections. 
These two kinds of uncertainties were combined to provide the total theoretical error of $\mathcal{B}^{\rm Exc/Inc}$, indicated by the subscript "${\rm th}$". The uncertainty due to the experimental error of the lifetimes is labeled by the subscript "$\tau$".

\textsc{Babar}'s recent measurements of the branching ratios of the radiative decays  $\tau \to l \bar{\nu} \nu \gamma$, with $l=e$ and $\mu$, for a minimum photon energy $\omega_0=10$~MeV in the $\tau$ rest frame, are~\cite{Lees:2015gea,OberhofPhDThesis}:
\begin{align}
   \mathcal{B}_{\scriptscriptstyle \rm EXP} \left(\tau \to e  \bar{\nu}  \nu \gamma  \right) 
   & \, = 1.847  \, (15)_{\rm st} (52)_{\rm sy} \times 10^{-2},
   \label{eqn:TAUEBabar} \\
  \mathcal{B}_{\scriptscriptstyle \rm EXP} \left(\tau \to \mu  \bar{\nu}  \nu \gamma  \right) 
  & \, = 3.69  \, (3)_{\rm st} (10)_{\rm sy} \times 10^{-3},
   \label{eqn:TAUMUBabar}
\end{align}
where the first error is statistical and the second is systematic. These results are substantially more precise than the previous measurements of the \textsc{Cleo} collaboration~\cite{Bergfeld:1999yh}. The experimental values in Eqs.~(\ref{eqn:TAUEBabar},\ref{eqn:TAUMUBabar}) were obtained requiring a signal with either a muon or an electron, plus a single photon; they must therefore be compared with our predictions for the exclusive branching ratios in table~\ref{tab:BR}. For $\tau \to \mu  \bar{\nu}  \nu \gamma$ decays, the branching ratio measurement and prediction agree within 1.1 standard deviations (1.1$\sigma$). On the contrary, the experimental and theoretical values for $\tau \to e \bar{\nu}  \nu \gamma$ decays differ by $2.02 \, (57) \times 10^{-3}$, i.e.\  by 3.5$\sigma$. If \textsc{Babar}'s measurement (\ref{eqn:TAUEBabar}) were inclusive, this discrepancy would decrease to 2.2$\sigma$. This puzzling discrepancy deserves further researches.

The branching ratio of radiative muon decays was measured long ago for a minimum photon energy $\omega_0=10$~MeV in the $\mu$ rest frame~\cite{Crittenden:1959hm}, and more recently by the \textsc{Meg} collaboration for $\omega_0=40$~MeV and minimum electron energy $E_e^{\rm min}=45$~MeV (in this case, $\mathcal{B}^{\rm Inc}$ and $\mathcal{B}^{\rm Exc}$ coincide):
\begin{align}
   &\mathcal{B}_{\scriptscriptstyle \rm EXP} \left(\mu \to e  \bar{\nu}  \nu \gamma,  \omega_0\!=\!10~\rm{MeV} \right) 
    \, = \, 1.4  \, (4) \times 10^{-2}~\cite{Crittenden:1959hm},
      \label{eqn:MUE1961} \\ 
   &\mathcal{B}_{\scriptscriptstyle \rm EXP} \left(\mu \to e  \bar{\nu}  \nu \gamma,  \omega_0\!=\!40~\rm{MeV}, 
   E_e^{\rm min}\!=\!45~\rm{MeV}  \right) 
    \, = \, 6.03  \, (14)_{\rm st} (53)_{\rm sy} \times 10^{-8}~\cite{Adam:2013gfn}.
   \label{eqn:MUEMEG}
\end{align}
Both measurements agree with our theoretical predictions (see table~\ref{tab:BR}). New precise results are expected from the \textsc{Meg}~\cite{Adam:2013gfn} and \textsc{Pibeta}~\cite{Pocanic:2015rrw} collaborations.

The relative magnitude of radiative corrections were also studied in the specific final-state configuration of the decays (\ref{eqn:muraddecay},\ref{eqn:tauraddecays}) when the neutrino energies ($E_\nu$ and $E_{\bar{\nu}}$) are very small, i.e.\ when the photon and the final charged lepton are almost back-to-back. As already mentioned in the introduction, this phase-space region is of particular interest for experiments searching for the charged lepton flavor violating decays $\mu \to e \gamma$ or $\tau \to l \gamma$. Indeed the SM decays (\ref{eqn:muraddecay},\ref{eqn:tauraddecays}) are indistinguishable from the signal ($\mu \to e \gamma$ or $\tau \to l \gamma$), except for the energy carried away by the neutrinos. This SM background can be suppressed via a precise determination of the final state momenta: the total energy of the $e \gamma$ final state (or $l\gamma$) must be as close as possible to $m_\mu$ (or $m_\tau$).

\begin{figure}[htbp]
\centering
\includegraphics[width=0.7\textwidth]{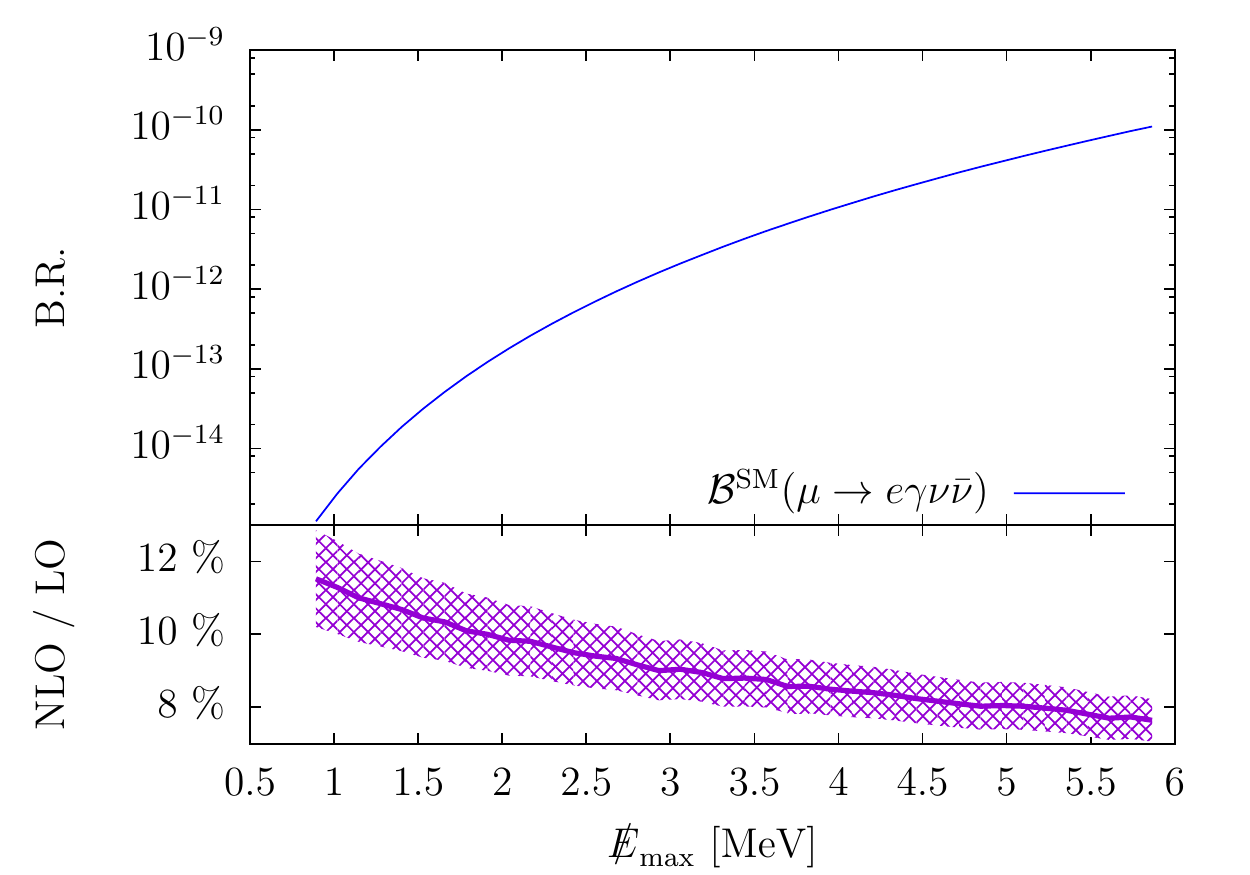}
\caption{Top panel: branching ratio of the radiative muon decay (2.1) as a function of the invisible energy cut $\slashed E_{\rm max}$. 
Lower panel: the ratio of NLO corrections with respect to the LO branching ratio. The purple band represents the assigned theoretical error due to uncomputed NNLO corrections.
}
\label{fig:br_inv_nlo} 
\end{figure}

The upper panel of figure~\ref{fig:br_inv_nlo} shows, for radiative muon decays, the SM prediction at NLO of the branching fraction 
$\mathcal{B}^{\mysmall SM} (\slashed E_{\rm max})$, 
defined as the integral of~\eqref{eq:radiativedecayrateunpolarizedNLO} over the phase space region satisfying $ \slashed E = E_\nu+E_{\bar{\nu}} = m_\mu-E_e - \omega \le \slashed E_{\rm max}$. The maximum missing energy $\slashed E_{\rm max}$ is assumed to be lower than the detection threshold $\omega_0$ and much lower than the muon mass: $\slashed E_{\rm max} < \omega_0 \ll m_\mu$. For this reason, $\slashed E_{\rm max}$ plays the role of infrared cut-off. Indeed if $\slashed E_{\rm max} \ll m_\mu$, the photon energy must be of the order of  $(m_\mu-\slashed E_{\rm max})/2$, a value much larger than the threshold $\omega_0$.
Moreover, at NLO also a second soft photon can be emitted, but its energy is always below the threshold $\omega_0$ -- and therefore invisible -- since it cannot exceed $\slashed E_{\rm max}$. We calculated and included these second-soft-photon effects in $\mathcal{B}^{\mysmall SM} (\slashed E_{\rm max})$ adopting the same $\ymin' \to 0$ limit described in~\cite{Fael:2015gua} for the numerical evaluation of the exclusive and inclusive branching fractions in table~\ref{tab:BR}. The lower panel of figure~\ref{fig:br_inv_nlo} shows the ratio of NLO corrections with respect to the LO branching ratio. The relative magnitude of these corrections can be as large as $8-12\%$ for an invisible energy cut $\slashed E_{\rm max}$ ranging from 1 to 6~MeV. The purple band represents the theoretical error assigned to this ratio for uncomputed NNLO corrections; it is estimated to be $\delta \mathcal{B}^{\mysmall SM}_{\mysmall NLO} (\slashed E_{\rm max}) \sim (\alpha/\pi) (\ln r) (\ln \frac{\slashed E_{\rm max}}{m_\mu}) \mathcal{B}^{\mysmall SM}_{\mysmall NLO} (\slashed E_{\rm max})$.

\section{$\tau$ dipole moments via radiative leptonic  $\tau$ decays}\label{sec:ad}

In this section we report on a recent proposal to determine the $\tau$ dipole moments via radiative leptonic $\tau$ decays~\cite{Fael:2013ij,FaelThesis,nextfael}. The most general vertex function describing the interaction between a photon and the initial and final states of an on-shell $\tau$ lepton can be written in the form
\begin{equation}
  \Gamma^\mu (q^2) = i e \left\{
  \gamma^\mu F_1(q^2)   
  + \frac{\sigma^{\mu\nu} q_\nu}{2m_\tau} \Big[ i F_2(q^2) + F_3 (q^2) \gamma_5 \Big] 
  + \left( \gamma^{\mu} - \frac{2m_{\tau}q^{\mu}}{q^2}\right) \! \gamma_5 \, F_4 (q^2) \right\},
  \label{eqn:ffgammavertex}
\end{equation}
where $e>0$ is the positron charge, $\sigma_{\mu\nu}=i\,[\gamma_\mu,\gamma_\nu]/2$, and $q$ is the ingoing four-momentum of the off-shell photon. In the limit $q^2 \to 0$,
$ F_{2}(0) = a_\tau$ and $F_{3}(0) = -d_\tau  (2m_\tau/e)$,
where $a_\tau = (g_{\tau}-2)/2$ and $d_\tau$ are the anomalous magnetic moment and EDM of the $\tau$, respectively.

Deviations of the $\tau$ dipole moments from the SM values can be analyzed in the framework of dimension-six gauge-invariant operators. Out of the complete set of 59 independent gauge invariant operators in~\cite{Buchmuller:1985jz,Grzadkowski:2010es}, only two of them can directly contribute to the $\tau$ \gmt and EDM at tree level (i.e., not through loop effects),
$Q^{33}_{lW} =  \left( \bar{l}_\tau \sigma^{\mu\nu} \tau_R \right) \sigma^I \varphi \,  W_{\mu\nu}^I$ and
$Q^{33}_{lB} = \left( \bar{l}_\tau \sigma^{\mu\nu} \tau_R \right) \varphi \, B_{\mu\nu}$,
where $\varphi$ and $l_\tau = (\nu_\tau,\tau_L)$ are the Higgs and the left-handed SU(2) doublets, $\sigma^I$ the Pauli matrices, and $W_{\mu\nu}^I$ and $B_{\mu\nu}$ are the gauge field strength tensors. The leading non-standard effective Lagrangian relevant for our study is therefore given by
\begin{equation}
 \mathcal{L}_{\rm eff} =
  \frac{1}{\Lambda^2}
  \left[C^{33}_{lW}Q^{33}_{lW}+ 
  C^{33}_{lB} Q^{33}_{lB} + {\rm h.c.} \right].
  \label{eqn:leff}
\end{equation}
After the electroweak symmetry breaking, the two operators mix and give additional, beyond the SM, contributions to the $\tau$ anomalous magnetic moment and EDM:
\begin{equation}
  \tilde{a}_\tau = \frac{2 m_\tau}{e} \frac{\sqrt{2} v}{\Lambda^2} 
  \,\, \Re \left[ \cw \Clb^{33} -\sw \Clw^{33} \right] ,\quad
  \tilde{d}_\tau =  \frac{\sqrt{2} v}{\Lambda^2} 
  \,\, \Im \left[ \cw \Clb^{33} -\sw \Clw^{33} \right],
\end{equation}
where $v=246$~GeV and $\sw$ is the weak mixing angle.

The present resolution on the $\tau$ anomalous magnetic moment is only of $\mathcal{O}(10^{-2})$, more than an order of magnitude larger than its precise SM prediction
$    a_{\tau}^{\mysmall \rm SM} = 
         117 \, 721 \, (5) \times 10^{-8}$~\cite{Eidelman:2007sb}.  
In fact, the short lifetime of the $\tau$ ($2.9\times 10^{-13}$~s) has so far prevented the determination of $a_{\tau}$ by measuring the $\tau$ spin precession in a magnetic field, like in the electron and muon $g$$-$$2$ experiments. 
The present PDG limit on the $\tau$ \gmt was derived by the \textsc{Delphi} collaboration from $e^+ e^- \to e^+ e^- \tau^+ \tau^-$ total cross section measurements at LEP2 (the study of $a_\tau$ via this channel was proposed in~\cite{Cornet:1995pw}). Limits were derived comparing these measurements with the SM values, assuming that possible deviations were due to non-SM contributions to $a_\tau$. The obtained limit at 95\% CL is~\cite{Abdallah:2003xd}
\begin{equation}
   -0.052 < \tilde{a}_\tau < 0.013.
   \label{eqn:atauexpbound95}
\end{equation}
The reanalysis of Ref.~\cite{GonzalezSprinberg:2000mk} of various LEP and SLD measurements -- mainly of the $e^+e^- \to \tau^+\tau^-$ cross section -- allowed the authors to set the indirect 2$\sigma$ confidence interval
\begin{equation}
   -0.007 < \tilde{a}_{\tau}  < 0.005, 
   \label{eqn:atauGonzalesbound}
\end{equation}
a bound stronger than that in~(\ref{eqn:atauexpbound95}). This analysis assumed $\tilde{d}_\tau = 0$. The bound~(\ref{eqn:atauGonzalesbound}) has been confirmed by a recent update in Ref.~\cite{nextfael}.

Lepton EDMs are predicted to be extremely small in the SM, of the $\mathcal{O}(10^{-38} - 10^{-35}) \, e\cdot$cm~\cite{Commins:1999jh}, far below the current experimental reach. The present PDG limit on the $\tau$ EDM at $95\%$~CL is
\begin{equation}
 - 2.2 < \mathrm{Re} (d_\tau) < 4.5 \; \; (10^{-17} \;  e \cdot \mathrm{cm}),
\quad
 - 2.5 < \mathrm{Im} (d_\tau) < 0.8 \; \; (10^{-17} \;  e \cdot \mathrm{cm}); 
\label{eq dtauexp}
\end{equation}
it was obtained by the Belle collaboration~\cite{Inami:2002ah} following the analysis of Ref.~\cite{Bernreuther:1993nd} for the impact of an effective operator for the $\tau$ EDM in the process $e^+ e^- \rightarrow \tau^+ \tau^-$.

At the LHC, bounds on the $\tau$ dipole moments are expected to be set in $\tau$ pair production via Drell-Yan~\cite{HayreterValencia} or double photon scattering processes~\cite{Atag:2010ja}. The best limits achievable with the former are estimated to be comparable with the current ones if one assumes that the total cross section for $\tau$ pair production will be measured at the $14\%$ level. 
Earlier proposals to set bounds on the $\tau$ dipole moments can be found in~\cite{Samuel:1992fm,delAguila:1991rm,EscribanoMasso}.

The Belle II experiment at the upcoming high-luminosity $B$ factory SuperKEKB~\cite{Aushev:2010bq} will offer new opportunities to improve the determination of the $\tau$ electromagnetic properties. The authors of Refs.~\cite{Bernabeu:2007rr,Bernabeu:2008ii} proposed to determine the Pauli form factor $F_{2}(q^2)$ of the $\tau$ via $\tau^+ \tau^-$ production in $e^+ e^-$ collisions at the $\Upsilon$ resonances ($\Upsilon$(1S), $\Upsilon$(2S) and $\Upsilon$(3S)) with a sensitivity of $O(10^{-5})$ or even better (of course, the center-of-mass energy at super $B$ factories is $\sqrt{s} \sim M_{\Upsilon(4S)} \approx 10$ GeV, so that the form factor $F_{2}(q^2)$ is not the anomalous magnetic moment). The contributions to the  $e^+e^- \to \tau^+ \tau^-$ cross section arise not only from the usual $s$-channel one-loop vertex corrections, but also from box diagrams, which should be somehow subtracted out. The strategy proposed in~\cite{Bernabeu:2007rr,Bernabeu:2008ii} to eliminate their contamination is to measure the observables on top of the $\Upsilon$ resonances, where the non-resonant box diagrams should be numerically negligible.
However, it is very difficult to resolve the narrow peaks of the $\Upsilon (1S,2S,3S)$ ($\Gamma_\Upsilon \sim$ 20--50~keV) in the $\tau^+ \tau^-$ decay channel (the $\Upsilon(4S)$ decays almost entirely in $B\bar{B}$) because of the natural irreducible beam energy spread associated to any $e^+ e^-$ synchrotron (5.45~MeV at SuperKEKB). In Ref.~\cite{nextfael} it was shown that, at the Belle II experiment, the $\tau^+ \tau^-$ events produced with beams at a center-of-mass energy $\sqrt{s} \sim M_\Upsilon$ will be mostly due to non-resonant contributions. The situation at Belle was similar (the energy spread at KEKB was 5.24~MeV). Therefore, the measurement of the $e^+ e^- \to \tau^+ \tau^-$ cross section on top of the $\Upsilon$ resonances will not eliminate the contamination of the non-resonant contributions.

The effective Lagrangian \eqref{eqn:leff} generates additional non-standard contributions to the differential decay rate in Eq.~\eqref{eqn:radiativedecayrateNLO}. For a $\tau^\pm$ decay they can be summarised in the shifts~\cite{nextfael}
\begin{equation}
	G  \,\to\,   G \,+\,  \tilde{a}_\tau \,  G_a, \quad
	J   \,\to\,   J  \,+\,  \tilde{a}_\tau \,  J_a, \quad
	K   \,\to\,   K \,+\,  \tilde{a}_\tau \,  K_a, \quad
	L   \,\to\,   L  \,\mp\,  \left(m_\tau/e \right) \,  \tilde{d}_\tau \, L_d.
	\label{eqn:Ldchange}
\end{equation}
Tiny terms of $O(\tilde{a}_{\tau}^2)$, $O(\tilde{d_{\tau}}^2)$ and $O(\tilde{a}_{\tau} \tilde{d_{\tau}})$ were neglected. 
Deviations of the $\tau$ dipole moments from the SM values can be determined comparing the SM prediction for the differential rate in Eq.~\eqref{eqn:radiativedecayrateNLO}, modified by the terms $G_a$, $J_a$, $K_a$ and $L_d$, with sufficiently precise data.

The possibility to set bounds on $\tilde{a}_{\tau}$ via radiative leptonic $\tau$ decays was suggested long ago in Ref.~\cite{Laursen:1983sm}. In that article the authors proposed to take advantage of a radiation zero of the differential decay rate in \eqref{eqn:radiativedecayrateNLO}. This zero occurs when, in the $\tau$ rest frame, the final lepton $l$ and the photon are back-to-back, and $l$ has maximal energy. Since a non-standard contribution to $a_{\tau}$ spoils this radiation zero, precise measurements of this phase-space region could be used to set bounds on its value. However, a Monte Carlo simulation in the conditions of the Belle experiment shows no significant improvement of the existing limits for $a_{\tau}$~\cite{nextfael}.

A more powerful method to extract $\tilde{a}_{\tau}$ and $\tilde{d}_{\tau}$ consists in the use of an unbinned maximum likelihood fit of events in the full phase space~\cite{nextfael}. In this approach, we considered $e^+e^- \to \tau^+ \tau^-$ events where both $\tau$ leptons decay subsequently into a particular final state: $\tau^{\mp}$ (signal side) decays to the radiative leptonic mode, and the other $\tau^{\pm}$ (tag side) decays to some well known mode with a large branching fraction. As a tag decay mode we chose $\tau^{\pm}\to\rho^{\pm}\nu\to\pi^{\pm}\pi^0\nu$, which also serves as spin analyser and allows us to be sensitive to the spin-dependent part of the differential decay width of the signal decay using effects of spin-spin correlation of the $\tau$ leptons~\cite{Tsai:1971vv}. With this technique we analyzed a data sample of $(\ell^{\mp}\nu\nu\gamma,~\pi^{\pm}\pi^0\nu)$ events corresponding to the total amount of data available at Belle and the one planned at the Belle II experiment.

The feasibility study of Ref.~\cite{nextfael} shows that the experimental sensitivity on $\tilde{a}_\tau$ that can be reached at the Belle II experiment can improve the \textsc{Delphi} bound~\eqref{eqn:atauexpbound95}. On the other hand, the expected sensitivity on the $\tau$ EDM is still worse than the most precise measurement of $\tilde{d}_\tau$ performed at Belle in $\tau$ pair production~\cite{Inami:2002ah}.

\section{Conclusions}\label{sec:c}

We discussed the SM predictions of the differential rates and branching ratios of the decays $\mu \to e \gamma \nu \bar{\nu}$ and $\tau \to l \gamma \nu \bar{\nu} \, (l=\mu,e)$ at NLO recently derived in Ref.~\cite{Fael:2015gua}. Our predictions agree with the measurements of the branching ratio $\mathcal{B} (\mu \to e \bar{\nu} \nu \gamma)$ obtained by \textsc{Meg} and in Ref.~\cite{Crittenden:1959hm}. Also the recent precise measurement by \textsc{Babar} of the branching ratio $\mathcal{B} (\tau \to \mu  \bar{\nu} \nu \gamma)$, for $\omega_0 = 10$~MeV, agrees with our prediction within 1.1 standard deviations (1.1$\sigma$). On the contrary, \textsc{Babar}'s recent measurement of the branching ratio $\mathcal{B} (\tau \to e \bar{\nu}  \nu \gamma)$, for the same threshold $\omega_0$, differs from our prediction by 3.5$\sigma$.  This puzzling discrepancy deserves further researches.

We proposed to determine the $\tau$ dipole moments via precise measurements of radiative leptonic $\tau$ decays at high-luminosity $B$ factories. Deviations of the $\tau$ \gmt and EDM from the SM predictions can be determined via an effective Lagrangian approach. Our dedicated feasibility study in Ref.~\cite{nextfael} showed that the measurement of the $\tau$ anomalous magnetic moment at the upcoming Belle~II experiment can improve the current bound of the \textsc{Delphi} experiment, while the foreseen sensitivity is not expected to lower the current limit on the $\tau$ EDM.

\acknowledgments

We would like to thank our colleagues S.~Eidelman, D.~Epifanov and L.~Mercolli for very useful discussions and correspondence.
The work of M.F.\ is supported by the Swiss National Science Foundation.
M.P.\ also thanks the Department of Physics and Astronomy of the University of Padova for its support. His work was supported in part by the Italian Ministero dell'Universit\`a e della Ricerca Scientifica under the program PRIN 2010-11, and by the European Program INVISIBLES (contract PITN-GA-2011-289442).




\begin{thebibliography}{50}

\bibitem{Michel:1949qe}
  L.~Michel,
  Proc.\ Phys.\ Soc.\ A {\bf 63}, 514 (1950).

\bibitem{Bouchiat:1957zz}
  C.~Bouchiat and L.~Michel,
  Phys.\ Rev.\  {\bf 106}, 170 (1957).
  
\bibitem{KinoshitaSirlin:1957}
  T.~Kinoshita and A.~Sirlin,
  Phys.\ Rev.\  {\bf 107}, 593 (1957); ibid.\  {\bf 108}, 844 (1957).
  
  
\bibitem{Eichenberger:1984gi}
  W.~Eichenberger, R.~Engfer and A.~Van Der Schaaf,
  Nucl.\ Phys.\ A {\bf 412}, 523 (1984).

\bibitem{Fetscher:1993ki}
  W.~Fetscher and H.~J.~Gerber,
  Adv.\ Ser.\ Direct.\ High Energy Phys.\  {\bf 14}, 657 (1995).

\bibitem{Adam:2013gfn} 
  J.~Adam {\it et al.} [MEG Collaboration],
  hep-ex/1312.3217.
  
\bibitem{Baldini:2013ke}
  A.~M.~Baldini {\it et al.},
   physics.ins-det/1301.7225.

\bibitem{Berger:2014vba}
  N.~Berger [Mu3e Collaboration],
  Nucl.\ Phys.\ Proc.\ Suppl.\  {\bf 248-250}, 35 (2014).
 
\bibitem{Flores-Tlalpa:2015vga}
  A.~Flores-Tlalpa, G.~L\'opez Castro and P.~Roig,
  hep-ph/1508.01822.
  
\bibitem{Lees:2015gea}
  J.~P.~Lees {\it et al.}  [BaBar Collaboration],
  Phys.\ Rev.\ D {\bf 91}, 051103 (2015)
  [hep-ex1502.01784].

\bibitem{OberhofPhDThesis}
  B.~Oberhof, {\em Measurement of $\mathcal{B}(\tau \to l \gamma \nu \bar{\nu}, l=e,\mu)$ at \textsc{Babar}}, 
  Ph.D.\ thesis, University of Pisa, 2015.

\bibitem{Fael:2015gua} 
  M.~Fael, L.~Mercolli and M.~Passera,
  JHEP {\bf 1507}, 153 (2015)
  [hep-ph/1506.03416].

\bibitem{Kinoshita:1958ru}
  T.~Kinoshita and A.~Sirlin,
  Phys.\ Rev.\  {\bf 113}, 1652 (1959).
  
\bibitem{Kinoshita:1962ur}
  T.~Kinoshita,
  J.\ Math.\ Phys.\  {\bf 3}, 650 (1962).

\bibitem{Lee:1964is}
  T.~D.~Lee and M.~Nauenberg,
  Phys.\ Rev.\  {\bf 133}, B1549 (1964).
    
\bibitem{Crittenden:1959hm}
  R.~R.~Crittenden, W.~D.~Walker and J.~Ballam,
  Phys.\ Rev.\  {\bf 121}, 1823 (1961).
  
  \bibitem{Pocanic:2015rrw}
  D.~Pocanic [PEN Collaboration],
  nucl-ex/1512.09355.
  
  \bibitem{Fael:2013ij}
  M.~Fael, L.~Mercolli and M.~Passera,
  Nucl.\ Phys.\ Proc.\ Suppl.\  {\bf 253-255}, 103 (2014)
  [hep-ph/1301.5302].
  
\bibitem{FaelThesis}
  M.~Fael, {\it Electromagnetic dipole moments of fermions},
  Ph.D.\ thesis, University of Padova, Italy \& University of Zurich, Switzerland, 2014;
  http://opac.nebis.ch/ediss/20142170.pdf.

\bibitem{nextfael}
  S.~Eidelman, D.~Epifanov, M.~Fael, L.~Mercolli and M.~Passera,
  hep-ph/1601.07987.
  
\bibitem{PDG2014}
  K.~A.~Olive {\it et al.}  [Particle Data Group Collaboration],
  Chin.\ Phys.\ C {\bf 38}, 090001 (2014).
  
\bibitem{AHKN}
  T.~Aoyama, M.~Hayakawa, T.~Kinoshita and M.~Nio,
  Phys.\ Rev.\ Lett.\  {\bf 109}, 111807 (2012)
  [hep-ph/1205.5368];
  Phys.\ Rev.\ D {\bf 91}, no.\ 3, 033006 (2015)
  [hep-ph/1412.8284].

\bibitem{Ferroglia:2013dga}
  A.~Ferroglia, C.~Greub, A.~Sirlin and Z.~Zhang,
  Phys.\ Rev.\ D {\bf 88}, no.\ 3,  033012 (2013)
  [hep-ph/1307.6900].
    
\bibitem{Fael:2013pja}
  M.~Fael, L.~Mercolli and M.~Passera,
  Phys.\ Rev.\ D {\bf 88}, no.\ 9,  093011 (2013)
  [hep-ph/1310.1081].

\bibitem{Fronsdal:1959zzb}
  C.~Fronsdal and H.~Uberall,
  Phys.\ Rev.\  {\bf 113}, 654 (1959).

\bibitem{EcksteinPratt}
  S.~G.~Eckstein and R.~H.~Pratt,
  Ann.\ Phys.\ {\bf 8}, 297 (1959).

\bibitem{Kuno:1999jp}
  Y.~Kuno and Y.~Okada,
  Rev.\ Mod.\ Phys.\  {\bf 73}, 151 (2001)
  [hep-ph/9909265].
    
\bibitem{Fischer:1994pn}
  A.~Fischer, T.~Kurosu and F.~Savatier,
  Phys.\ Rev.\ D {\bf 49}, 3426 (1994).
 
\bibitem{Arbuzov:2004wr}
  A.~B.~Arbuzov and E.~S.~Scherbakova,
  Phys.\ Lett.\ B {\bf 597}, 285 (2004)
  [hep-ph/0404094].

\bibitem{KSPRL1959}
 T.~Kinoshita and A.~Sirlin, 
 Phys.\ Rev.\ Lett.\ \textbf{2}, 177 (1959).

\bibitem{Bergfeld:1999yh} 
  T.~Bergfeld {\it et al.} [CLEO Collaboration],
  Phys.\ Rev.\ Lett.\  {\bf 84}, 830 (2000)
  [hep-ex/9909050].

\bibitem{Buchmuller:1985jz} 
  W.~Buchmuller and D.~Wyler,
  Nucl.\ Phys.\ B {\bf 268}, 621 (1986).

\bibitem{Grzadkowski:2010es} 
  B.~Grzadkowski, M.~Iskrzynski, M.~Misiak and J.~Rosiek,
  JHEP {\bf 1010}, 085 (2010)
  [hep-ph/1008.4884].

\bibitem{Eidelman:2007sb} 
  S.~Eidelman and M.~Passera,
  Mod.\ Phys.\ Lett.\ A {\bf 22}, 159 (2007) 
  [hep-ph/0701260].
  
 \bibitem{Cornet:1995pw} 
  F.~Cornet and J.~I.~Illana,
  Phys.\ Rev.\ D {\bf 53}, 1181 (1996)
  [hep-ph/9503466].  

 \bibitem{Abdallah:2003xd} 
  J.~Abdallah {\it et al.} [DELPHI Collaboration],
  Eur.\ Phys.\ J.\ C {\bf 35}, 159 (2004)
  [hep-ex/0406010].   

 \bibitem{GonzalezSprinberg:2000mk} 
  G.~A.~Gonzalez-Sprinberg, A.~Santamaria, J.~Vidal,
  Nucl.\ Phys.\ B {\bf 582}, 3 (2000)
  [hep-ph/0002203].

 \bibitem{Commins:1999jh} 
  E.~D.~Commins,
  Adv.\ At.\ Mol.\ Opt.\ Phys.\  {\bf 40}, 1 (1999). 

  \bibitem{Inami:2002ah} 
  K.~Inami {\it et al.} [Belle Collaboration],
  Phys.\ Lett.\ B {\bf 551}, 16 (2003)
  [hep-ex/0210066].

  \bibitem{Bernreuther:1993nd} 
  W.~Bernreuther, O.~Nachtmann and P.~Overmann,
  Phys.\ Rev.\ D {\bf 48}, 78 (1993).

  \bibitem{HayreterValencia} 
  A.~Hayreter and G.~Valencia,
  Phys.\ Rev.\ D {\bf 88}, no. 1, 013015 (2013)
  [ibid.\ {\bf 91}, no. 9, 099902 (2015) Erratum]
  [hep-ph/1305.6833];
  JHEP {\bf 1507}, 174 (2015)
  [hep-ph/1505.02176].
  
  \bibitem{Atag:2010ja} 
  S.~Atag and A.~A.~Billur,
  JHEP {\bf 1011}, 060 (2010)
  [hep-ph/1005.2841].

\bibitem{Samuel:1992fm} 
  M.~A.~Samuel and G.~Li,
  Int.\ J.\ Theor.\ Phys.\  {\bf 33}, 1471 (1994).

  \bibitem{delAguila:1991rm} 
  F.~del Aguila, F.~Cornet and J.~I.~Illana,
  Phys.\ Lett.\ B {\bf 271}, 256 (1991).

\bibitem{EscribanoMasso} 
  R.~Escribano and E.~Masso,
  Phys.\ Lett.\ B {\bf 301}, 419 (1993);
  ibid.\ {\bf 395}, 369 (1997)
  [hep-ph/9609423].
  
 \bibitem{Aushev:2010bq} 
  T.~Aushev {\it et al.},
  hep-ex/1002.5012.

\bibitem{Bernabeu:2007rr} 
  J.~Bernabeu, G.~A.~Gonzalez-Sprinberg, J.~Papavassiliou and J.~Vidal,
  Nucl.\ Phys.\ B {\bf 790}, 160 (2008)
  [hep-ph/0707.2496].
  
\bibitem{Bernabeu:2008ii} 
  J.~Bernabeu, G.~A.~Gonzalez-Sprinberg, J.~Vidal,
  JHEP {\bf 0901}, 062 (2009)
  [hep-ph/0807.2366].
  
\bibitem{Laursen:1983sm} 
  M.~L.~Laursen, M.~A.~Samuel,  A.~Sen,
  Phys.\ Rev.\ D {\bf 29}, 2652 (1984)
  [ibid.\ {\bf 56}, 3155 (1997) Erratum]. 
  
  \bibitem{Tsai:1971vv} 
  Y.~S.~Tsai,
  Phys.\ Rev.\ D {\bf 4}, 2821 (1971)
  [ibid.\ {\bf 13}, 771 (1976) Erratum].
  
  
\end{thebibliography}
\end{document}